\documentstyle[aps,prl,twocolumn,psfig]{revtex}

\begin{document}
\draft                       


\title{ Molecular dynamics calculation of the thermal conductivity \\
       of vitreous silica}

\author{Philippe Jund and R\'emi Jullien}

\address{Laboratoire des Verres - Universit\'e Montpellier 2\\
         Place E. Bataillon Case 069, 34095 Montpellier France
	}


\maketitle


\begin{abstract}
We use extensive classical molecular dynamics simulations to calculate
the thermal conductivity of a model silica glass. Apart from the
potential parameters, this is done with no other adjustable quantity
and the standard equations of heat transport are used directly
in the simulation box. The calculations have been done 
between 10 and 1000 Kelvin and the results are in good 
agreement with the experimental data at temperatures above 20K. 
The plateau observed around 10K can be accounted for by correcting 
our results taking into account finite size effects in a 
phenomenological way.
\end{abstract}
 


\narrowtext

\section{Introduction}

The thermal properties of glasses exhibit some specific and unusual 
features which are well known for quite some time \cite{bruck}. These
features are apparent in the specific heat and the thermal conductivity but
we would like to focus here on the thermal conductivity $\kappa$. The temperature dependence
of $\kappa (T)$ can be separated in 3 distinct temperature domains:\\
- At very low temperature ($T \leq 1$K) the thermal conductivity increases
like $T^2$. This increase can be explained within the tunneling model \cite{tunnel} which has been proposed almost thirty years ago.\\
- At intermediate temperatures ($ 2 \leq T \leq 20$K) the thermal conductivity
exhibits a ``plateau'' for which several explanations have been given 
\cite{plateau}. An extension of the tunneling model, the
soft-potential model, has been proposed and gives a coherent description
of the plateau by introducing the concept of ``soft vibrations'' \cite{sp1,sp2}.\\
- At high temperature, ($T \geq 30$K), $\kappa(T)$ rises smoothly and seems to saturate to a
limiting value $\kappa_{\infty}$ unlike crystals where $\kappa(T) \sim 1/T$ at
elevated temperature. Recently this second rise of the thermal conductivity
has also been explained within the soft-potential model \cite{gil} which appears
to be able to account for all the thermal anomalies of glasses over
the whole temperature range.\\
Our aim here is not to propose a new or alternative explanation of the
above mentioned anomalies. The purpose is to perform a molecular
dynamics (MD) simulation on a model silica glass using a very widely
used interaction potential (the so-called ``BKS'' potential \cite{bks}) 
without any pre-conception of the model able to explain the thermal 
anomalies of silica. This means that we do not add or inject an 
{\em a priori} quantity in the potential to reproduce a specific model.
We use the standard definition of the heat transport coefficients 
that we calculate directly in our simulation box. 
In fact we introduce artificially inside the system a ``hot'' and a ``cold'' 
plate which therefore induce a heat flux. This flux creates a temperature 
gradient and once the steady state has been reached we can determine 
the thermal conductivity. By using plates compatible with the periodic
boundary conditions we are able to calculate the thermal conductivity
directly during the simulations without any additional parameter. 
This technique has been inspired by earlier studies \cite{tenenbaum} in which
the plates were treated like hard walls and has mainly been applied to
the calculation of the thermal conductivity in 1- or 
2-dimensional systems \cite{maeda,micha}. Nevertheless very recently 
Oligschleger and Sch$\ddot{o}$n applied the same method in a study of 
heat transport phenomena in crystalline and glassy samples (mainly selenium) 
\cite{olig}. In parallel to these studies which can be 
called {\em in situ}, other
methods relying on the use of the density and heat flux correlation
functions \cite{ladd} or on the Kubo and Greenwood-Kubo formalism \cite{allen}
have been developed in order to determine the thermal conductivity of solids.
Our results for the thermal conductivity obtained with the BKS potential 
compare reasonably well with the experimental data. First of all the order of 
magnitude is correct above 20K and, at least in the range 20K-400K, a nice 
quantitative agreement is obtained.
Furthermore, by taking care of finite-size corrections in a very simple
phenomenological way, we are able to reproduce the plateau around 10K.
Of course, the very low temperature $T^2$ behavior, which is known to
be due to quantum effects, is out of the scope of such a classical 
calculation.\\
This paper is organized in the following way. In section II we describe the
{\em modus operandi} we have used to obtain the thermal conductivity.
In section III we present first the results obtained directly from the
MD simulations. Then we show the effect of finite-size corrections on
these results and discuss our findings. In section IV we draw the major 
conclusions.

\section{Modus Operandi}

Except the determination of $\kappa(T)$, the simulations are standard
classical MD calculations on a microcanonical ensemble of 648 
particles (216 SiO$_2$ molecules) interacting via the BKS potential. Like
in a previous study \cite{pma} the particles are packed in a cubic box of 
edge length $L=21.48$\AA\ (the density is approximately equal to 2.18g/cm$^3$)
on which periodic boundary conditions are applied to simulate a macroscopic 
sample. The equations of motion are integrated using a fourth order
Runge-Kutta algorithm with a time step $\Delta t$ equal to $0.7$fs.
The glassy samples are obtained after a quench
from the liquid state ($T \approx 7000$K) at a constant quenching rate
of $2.3 \times 10^{14}$K/s.\\
The principle of the thermal conductivity determination is illustrated
in Fig.1. We consider two plates $P_-$ and $P_+$ perpendicular to the Ox axis 
and located at $x=-L/4$ and $x=+L/4$. These plates have a width $2\delta$ along
Ox and their surface is $L^2$. The positions of these plates permit to keep
the periodic boundary conditions without introducing an asymmetry in the
system. This has the advantage, compared to other studies \cite{kabu} in which
the introduction of the thermostatic plates breaks the symmetry, to
use a relatively small number of particles. At each iteration the 
particles which are inside $P_-$ and $P_+$ are
determined and their number is respectively $N_-$ and $N_+$. Once these
particles  are determined a constant energy $\Delta\epsilon$ is subtracted
from the energy of the particles inside $P_-$ and added to the energy 
of the particles in $P_+$.
 By imposing the heat transfer in this manner we insure a constant 
heat flux per unit area $J_x$ \cite{reif} which is equal 
to $\Delta\epsilon/(2L^2\Delta t)$. 
(the factor 2 comes from the fact that the heat flux coming from the
hot plate splits equally into two parts to reach the cold plate).
The energy modification is done by rescaling the velocities of the 
particles inside the plates. Nevertheless to avoid an artificial drift of 
the kinetic energy this has to be done with the total momentum of the plates 
being conserved. For a particle $i$ inside $P_-$ or $P_+$ the modified 
velocity is given at each iteration by
\begin{equation}
\vec{v_i}' = \vec{v_G} + \alpha(\vec{v_i} - \vec{v_G})
\end{equation}
where $\vec{v_G}$ is the velocity of the center of mass of the ensemble 
of particles in the plate and
\begin{equation}
\alpha = \sqrt{1 \pm \frac{\Delta\epsilon}{E_c^R}}
\end{equation}
depending on whether the particles are inside $P_+$ or $P_-$. 
The relative kinetic energy $E_c^R$ is given by
\begin{equation}
E_c^R = \frac{1}{2}\sum_i m_i\vec{v_i}^2 - \frac{1}{2}\sum_i m_i\vec{v_G}^2
\end{equation}
  
Following the standard definition of the transport coefficients \cite{reif}
the thermal conductivity is given by
\begin{equation}
\kappa = - \frac{J_x}{\partial T/ \partial x}
\end{equation}
where $\partial T / \partial x$ is the temperature gradient along Ox.
This formula, known as the Fourier's law of heat flow, is only valid
when a stable, linear temperature profile is obtained in the system. 
To calculate the gradient we divide the simulation box into $N_s$ ``slices''
along Ox in which the temperature is calculated at each iteration. Due to
the periodic boundary conditions we can concentrate only on the $N_s/2$ slices 
between $x=-L/4$ and $x=L/4$ and have a better determination of the temperature
in these slices since by symmetry arguments these slices are equivalent to
the $N_s/2$ slices located outside $[-L/4,L/4]$. We can  therefore determine
the temperature $T_i$ ($i= 1, ..N_s/2$)
of each slice  at each iteration. By averaging each $T_i$ over a large number 
of iterations to kill the unavoidable large temperature fluctuations 
(due to the small average number of particles in each slice), we are able to
determine after which simulation time $\tau$ the averaged profile
of $T(x)$ can reasonably well be approximated by a straight line. 
After that time we 
estimate $T(x)$ using a first order least square fit of the 
averaged $T_i$'s, the slope of which will give us the temperature gradient.    
At that point all the quantities necessary to calculate $\kappa$ are
determined. \\
Concerning the ``practical details'' of the simulation we have checked that
the results are independent on the choice of $\Delta\epsilon$ and for 
the other quantities we have used a compromise between computer time and 
accuracy of the results.
Here are the values used in our simulations: the width of the plates has
been taken equal to $2\delta = 1$\AA\ which means that approximately 30-40
atoms are inside the plates at each iteration. The temperature gradient has 
been determined on $N_s/2 =6$ slices, each slice containing approximately
100 particles. $\kappa$ has been determined on samples which have been 
saved all along the quenching procedure and therefore have different
temperatures T. To have the same treatment for each sample we have
fixed $\Delta\epsilon$ to 1\% of $k_BT$ which appears to be a good choice.
The temperature gradients obtained this way are small enough to insure
the validity of Eq. 4. The most problematic choice is the 
simulation time $\tau$. Indeed in order
to reach the steady state one needs long MD runs. For us a typical
run consists of 50000 MD steps (35 ps) directly after the quench 
during which the average temperature is fixed and the heat transfer is
switched on. Then we perform 450000 supplemental steps (315 ps) with only the
heat transfer but no other constraints during which the results are 
collected and averaged. After this time the temperature 
gradient should have converged and the value of $\kappa$ should be constant.
As we can see in figure 2,  this can be considered to be qualitatively 
true for the samples above  10K but certainly not for the low temperature
systems. In fact at low temperature longer runs (1 million
steps (700 ps)) are necessary and still the convergence is not perfect (it is
interesting to note that though our method converges slowly, it still
converges faster than the calculation of $\kappa(t)$ given by a steady state
experiment without a temperature gradient (\cite{tildes}, p.61)).
It is also worth noticing that the characteristic sigmoidal shape of the
temperature profile observed at 1K is consistent to  what is expected in the
intermediate regime where only heat transport over a small distance 
close to the plates is effective. In the following, only the results above 8K will be reported.

\section{Results}

The results obtained for the thermal conductivity as a function of
temperature in our model silica glass are reproduced in Fig.3 and compared 
to experimental data collected between 1 and 100K \cite{steph} and up
to 1000K \cite{zarzy}. The first observation is that                              our simulations with the BKS potential give the correct order of magnitude
over the whole temperature range  (except at very low temperatures)
 with no adjustable parameters apart from
the ``technical parameters'' described above and the constitutive potential
parameters. At very high temperatures, say above 500K, one observes
a more marked saturation of $\kappa(T)$ than in the experiments. 
This might be explained by the fact that other contributions than the one
described here can occur in the experiments at such high temperatures.
It is known that the radiative contributions (photon transport) in particular
increase quickly in this temperature range and can become of the order
of the phonon contributions \cite{zarzy}. 
In a large intermediate range, 20K to 400K, the agreement
between the calculated and experimental values is very good. Indeed in the
simulation also $\kappa$ increases in this temperature range unlike what is
found in crystalline samples. The major discrepancy between the simulation and 
the experiment can be seen between 8  and 20K since  we do not find
the characteristic plateau in the thermal conductivity.
In the following, we would like to argue that this discrepancy is essentially 
due to finite size effects.\\
In our cubic finite simulation box with periodic boundary
conditions, the components of the $\overrightarrow k$ wavevectors take
discrete values of the form $k_x = n_x 2\pi/L$, where $n_x$ is a 
relative integer (and similarly for the other space directions), and one 
cannot find, in principle, propagative
phonons with a frequency smaller than a lower cut-off $\omega_c$ which can
be estimated by $2\pi v_T/L$, where $v_T$ is the transverse sound velocity.
Considering the experimental value $v_T = 3.75\times 10^5$ cm/s for silica\cite{velo} this gives
$\omega_c/2\pi \simeq 1.5$ THz (in practice, when diagonalizing the dynamical
matrix in our low temperature sample, we find, similarly to a previous work done on the same system \cite{spectrum}, a slightly lower first non-zero
frequency $\omega_o/2\pi \simeq 1.2$ THz, in agreement with the existence of 
an excess of modes (maybe non-propagative), the so-called Boson peak\cite{boson}, in this frequency range \cite{discuss}.      
Therefore using the correspondence $\hbar\omega =  3k_BT$ which gives the
average phonon frequency $\omega$ of the phonons excited at temperature $T$,
there are certainly not enough phonons excited at temperatures below $T_o\simeq 19K$ in our box to be able to reproduce the experimental curve correctly. 
In Fig.3, the departure between our simulations and experiments is actually
seen at a temperature of the order of 20K, in good agreement with this 
analysis.\\
To try to put this argument on more quantitative grounds, let us assume that the thermal conductivity is given by the usual formula \cite{kit},
\begin{equation}
\kappa = {1\over 3}Cv\ell
\end{equation}
where $C$ is the heat capacity per unit volume, $v$ and $\ell$ the velocity 
and mean free path of the phonons, respectively. When applying such a
formula to glasses one has to be careful because of localization effects. 
Obviously $v$ and $\ell$ are the characteristics of the ``propagative''
 phonons, i.e. those which really contribute to the transport phenomena.
Consequently the heat capacity $C$ to be considered should be only due to
the contribution of these phonons and therefore (according to 
other authors \cite{tunnel,sp1}) should exhibit at low temperature
the usual Debye behavior (the same as in crystals). 
If we assume also that the lack of phonons in our box, i.e. a wrong value 
of $C$, is the essential cause for the underestimated calculated 
value of $\kappa$, a very simple and crude way to take care of this 
is to multiply our simulation results by a corrective factor 
$C_\infty/C_b$ which can be estimated
by taking for $C_\infty$ and $C_b$ the heat capacities calculated in the Debye
approximation for an infinite system and a finite cubic box of edge $L$, respectively.
To calculate this temperature dependent factor we have used the standard formulae\cite{kit}
\begin{equation}
C_\infty = {k_B\over 2\pi^2} ({1\over v_L^3} + {2\over v_T^3})\int_0^{\omega_D}
\Bigl({\hbar\omega/2k_BT \over\sinh(\hbar\omega/2k_BT)}\Bigr)^2\omega^2d\omega
\end{equation}
\begin{equation}
C_b = {k_B\over L^3} \sum_p\sum_{\overrightarrow k}\Bigl({\hbar v_pk/2k_BT 
\over\sinh(\hbar v_pk/2k_BT)}\Bigr)^2
\end{equation}
with $\omega_D^3 = (N/L^3)18\pi^2\bigl(1/v_L^3+2/v_T^3\bigr)^{-1}$.
In the expression of $C_b$ the double sum runs over the three polarizations $p=L,T_1,T_2$
and over the  first 
$N$ $\overrightarrow k$ vectors (quantized as indicated above) of 
lowest norm 
$k=|\overrightarrow k|$. For $N$ and $L$ we have taken the simulation values $N=648$ and $L=21.48$\AA\ and for the sound
velocities the experimental values $v_L = 5.9\times 10^5$ cm/s and 
$v_{T_1} = v_{T_2} = 3.75\times 10^5$ cm/s \cite{velo}.
When correcting our numerical data this way, we obtain the open squares 
represented in Fig.3 which turn out to be in very good agreement with 
the experimental results in the plateau region.
Of course, our reasoning is very crude since it assumes that 
finite size corrections affect only the heat capacity contribution in the
expression of $\kappa$ (Eq.5) and that the harmonic approximation 
holds for the propagative phonons  in that temperature range, however we think
that the agreement with the data cannot be fortuitous.                                            
It is unfortunate that we could not obtain more reliable results at 
temperatures lower than 8K (due to the impossibility to reach the permanent regime). Anyway, after correction, these results would
certainly give larger values for $\kappa$ than the experiments since it is 
known that, at very low temperatures, the propagative phonons start to be 
scattered on the quantum two level systems\cite{tunnel}
and therefore should have a lower mean free path than the one obtained 
in a classical calculation like the one performed here.

\section{Conclusion}

In conclusion we have presented the results of an extensive classical 
molecular dynamics simulation aimed to determine the thermal 
conductivity in a model silica glass. This determination has been done
directly inside the MD scheme with the use of the standard equations
governing the macroscopic transport coefficients and no pre-conceived model has
been assumed. Moreover it turns out that this method has considerable 
advantages (especially concerning the length of the simulations) compared 
to the standard methods usually implemented to 
calculate the transport coefficients \cite{tildes}.
The calculated values of the thermal conductivity are in good agreement 
with the experimental data at high temperature ($T > 20$K) and by
including finite size corrections in a simple way we are able to reproduce 
the plateau in the thermal conductivity around $10$K,
which has been the topic of several interpretations 
in the literature \cite{plateau}. The agreement between the calculated
and the experimental values of the thermal conductivity is even more 
striking when taking into account the ultra-fast quenching rate used 
to generate our amorphous samples. This shows once more the good quality
of the BKS potential which permits to reproduce the thermal anomalies
of vitreous silica with no additional parameters.\\  
Of course, our arguments on the finite size effects should be tested in 
the future by running larger samples. Nevertheless the simple phenomenological 
correction is so efficient that one can reasonably claim that the initial
discrepancy between the calculated and experimental values of the thermal
conductivity is indeed due to finite size effects and not to a weakness
of the method. Therefore we believe that this technique is a good way
to calculate the thermal properties of materials directly inside 
molecular dynamics simulations.\\
Most of the numerical calculations have been done on the IBM/SP2 
computer at CNUSC (Centre National Universitaire Sud de Calcul), Montpellier.
We would like to thank Claire Levelut and Jacques Pelous for very 
interesting comments.


%

\begin{figure}
\psfig{file=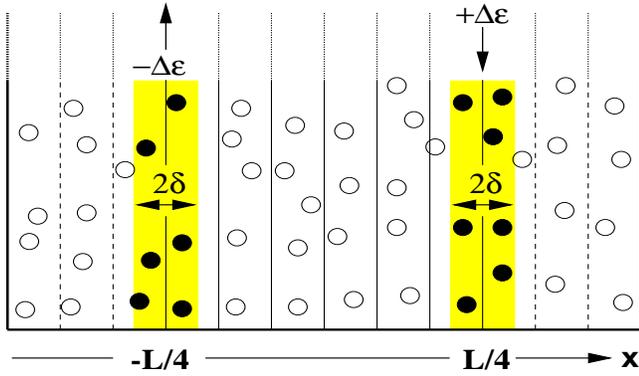,width=8.5cm,height=5cm}
\caption{
Schematic representation of the method used to determine the thermal
conductivity. More details can be found in the text. 
}
\label{Fig1}
\end{figure}

\vspace*{-20mm}
\begin{figure}
\psfig{file=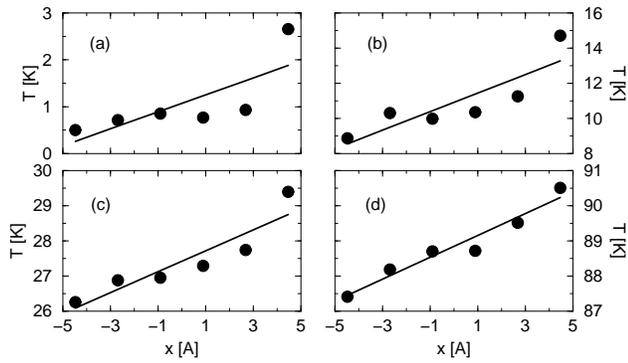,width=8.5cm,height=7cm}
\caption{
Values of the temperature as a function of $x$ in the slices
located between $x=-L/4$ and $x=L/4$ for 4 different samples and the
corresponding least square linear fit: \\
(a) $T \approx 1$K; (b)
$T \approx 11$K; (c) $T \approx 27$K and (d) $T \approx 89$K.
}
\label{Fig2}
\end{figure}

\vspace*{-20mm}
\begin{figure}
\psfig{file=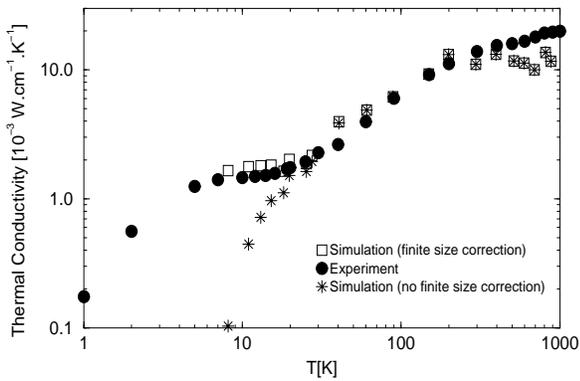,width=8.5cm,height=7.5cm}
\caption{
Log-log plot of the thermal conductivity as a function of temperature
in silica: \\
$\bullet$: experiment; $\ast$: simulations; $\Box$:
simulations with finite-size corrections.
}
\label{Fig3}
\end{figure}



\begin{references}
\bibitem{discuss} We do find (like other authors\cite{spectrum})
 an excess of modes at low frequencies compared with the Debye formula 
when diagonalizing the dynamical matrix. However the maximum of this excess
of modes is located at a frequency of about 2THz, i.e. almost twice 
the experimental value. We think that this discrepancy is 
simply due to finite size effects : by reducing the lowest frequency 
$\omega_o$ one would shift this maximum excess of modes to lower frequencies.
\end{references}
\end{document}